\documentclass{aastex63}

\newcommand{\revision}{}
\newcommand{\finrevision}{}

\submitjournal{ApJ}

\shorttitle{An Energy Balance Model for Rapidly and Synchronously Rotating Terrestrial Planets}
\shortauthors{Haqq-Misra and Hayworth}
\graphicspath{{./}{figures/}}

\begin{document}

\title{An Energy Balance Model for Rapidly and Synchronously Rotating Terrestrial Planets}

\correspondingauthor{Jacob Haqq-Misra}
\email{jacob@bmsis.org}
\author[0000-0003-4346-2611]{Jacob Haqq-Misra}
\affiliation{Blue Marble Space Institute of Science, Seattle, WA, USA}
\author[0000-0001-7132-035X]{Benjamin P. C. Hayworth}
\affiliation{Pennsylvania State University, University Park, PA, USA}



\begin{abstract}

This paper describes the Habitable Energy balance model for eXoplaneT ObseRvations (HEXTOR), which is a model for calculating latitudinal temperature profiles on Earth and other rapidly rotating planets. HEXTOR includes a lookup table method for calculating the outgoing infrared radiative flux and planetary albedo, which provides improvements over other approaches at parameterizing radiative transfer in an energy balance model. Validation cases are presented for present-day Earth and other Earth-sized planets with aquaplanet and land planet conditions from 0 to 45 degrees obliquity. A \revision{tidally locked} coordinate \revision{system} is also \revision{implemented in} the energy balance model, which enables calculation of the \revision{horizontal} temperature profile for planets in synchronous rotation around low mass stars. This coordinate transformed model is applied to cases for TRAPPIST-1e as defined by the TRAPPIST Habitable Atmosphere Intercomparison protocol, which demonstrates better agreement with general circulation models compared to the latitudinal energy balance model. Advances in applying energy balance models to exoplanets can be made by using general circulation models as a benchmark for tuning as well as by conducting intercomparisions between energy balance models with different physical parameterizations.

\end{abstract}



\section{Introduction} \label{sec:intro}

\revision{Space-based and ground-based surveys continue to reveal that most stars in the galaxy host planets \citep[e.g.][]{cassan2012one, dressing2015occurrence}. Statistical analyses of these results have suggested that Earth-sized planets capable of sustaining surface liquid water---known as habitable planets---should also be prevalent in up to half of G-, K-, and M-dwarf systems \citep[e.g.][]{dressing2015occurrence,Thompson2018, Hardegree-Ullman2019,Bryson2020note,bryson2021}. The detection of habitable Earth-sized exoplanets remains an ongoing priority for exoplanet science, with the goal of spectroscopically characterizing the atmospheres of such planets. This goal was emphasized in the 2020 Astrophysics Decadal Survey report \citep{NationalAcademiesofSciencesEngineering2021}, which recommended that NASA develop a $\sim$6 meter infrared/optical/ultraviolet flagship mission ``to search for biosignatures from a robust
number of about $\sim$25 habitable zone planets and to be a transformative facility for general astrophysics.'' Such a mission is not expected to launch until the 2040's, but the characterization of exoplanet atmospheres, and the search for remotely detectable biosignatures as evidence of life, will continue to advance through facilities such as TESS, JWST, PLATO, Extremely Large Telescopes (ELTs), and others.}

\revision{Computational climate models provide theoretical constraints on exoplanet habitability and are an important tool in the preparation for and analysis of such observations. Simplified climate models have historically provided constraints on the circumstellar habitable zone for exoplanetary systems \citep[e.g.][]{kasting1993habitable,kopparapu2013habitable,Kopparapu2014}, which describe the orbital limits where a terrestrial planet with a long-term carbon cycle (i.e., a carbonate-silicate cycle) could maintain liquid water on its surface. Numerous studies with more complex models have similarly been used to examine the plausible climates of known and theoretical exoplanetary systems, with the goal of understanding the limits of habitability in preparation for future observations \citep[see e.g.,][]{pierrehumbert2010principles,seager2010exoplanet,catling2017atmospheric,perryman2018exoplanet}. Climate models of varying complexity can be used and compared with one another to understand the physical processes that operate on other planets.}

Energy balance models (EBMs) describe the changes in a planet's surface temperature based on the balance between incoming radiation from the host star and outgoing infrared radiation from the planet. Computational EBMs were first used by \citet{budyko1969effect} and \citet{sellers1969global} \revision{to demonstrate that Earth's climate exhibits hysteresis between cold and warm states. Since these initial experiments, EBMs} have been widely used for understanding Earth's present and past climate (see \citet{north2017energy} for a review). The discovery of exoplanets has also motivated numerous EBM studies that explore plausible climates of other terrestrial and giant planets to aid in ongoing attempts at \revision{identifying habitable planets. This has included investigations of: 
the extent to which of rotation rate \citep{spiegel2008habitable,bahraminasr2020evaluating}, obliquity \citep{williams1997habitable,spiegel2009habitable,armstrong2014effects,rose2017ice,bahraminasr2020evaluating,palubski2020habitability}, eccentricity \citep{dressing2010habitable,palubski2020habitability}, and Milankovitch cycles \citep{spiegel2010generalized,haqq2014damping,forgan2016milankovitch,deitrick2018exo,deitrick2018exo2,quarles2021milankovitch} can affect habitability;
the effects of varying surface albedo \citep{shields2013effect,rushby2019effect,bahraminasr2020evaluating} and surface pressure \citep{vladilo2013habitable,ramirez2020effect,bahraminasr2020evaluating} on climate;
the role of mantle degassing and the carbonate-silicate cycle on climate evolution \citep{kadoya2014conditions,menou2015climate,kadoya2015evolutionary,haqq2016limit,kadoya2016evolutionary,kadoya2019outer};
the climate stability of synchronously rotating planets around low-mass stars \citep{kite2011climate,checlair2017no};
the possibility of habitable exomoons \citep{forgan2014effect,forgan2016exomoon}; 
and plausible climate for planets in binary star systems \citep{forgan2014assessing,forgan2016milankovitch,may2016examining,haqq2019constraining,okuya2019effects,yadavalli2020effects,quarles2021milankovitch}.}

This paper describes the Habitable EBM for eXoplaneT ObseRvations (HEXTOR), which was originally developed by \citet{williams1997habitable} and has subsequently been updated for use in a range of studies for \revision{Earth \citep{haqq2014damping}, Mars \citep{fairen2012reduced,batalha2016climate,hayworth2020warming}, and exoplanets \citep{haqq2016limit,haqq2019constraining}.} HEXTOR includes a newly improved lookup table method for calculating the outgoing infrared radiative flux and planetary albedo, which provides improvements over linear and polynomial representations that have been used in previous versions of the model. HEXTOR is able to reproduce the mean longitudinal temperature profile of present-day Earth and can be used to investigate the steady-state and temporal evolution of temperature for rapidly rotating terrestrial planets. HEXTOR also \revision{implements a tidally-locked coordinate system} for planets in synchronous rotation around low mass stars \revision{\citep{fortney2010transmission,koll2015deciphering,checlair2017no}}. This provides a mechanism to represent the warm day side and cool night side of such planets. This \revision{tidally-locked coordinate system} is applied to the four cases from the TRAPPIST Habitable Atmosphere Intercomparison (THAI), which was initially designed to compare benchmark climate states for TRAPPIST-1e between four general circulation models (GCMs). The comparison of \revision{tidally-locked} EBM results with GCM results for the THAI cases highlights areas for synergy between one-dimensional and higher dimensional models.

\section{Parameterizing Radiative Transfer} \label{sec:EBMdescription}

The energy balance equation for a terrestrial planet relates the change in surface temperature,  $dT/dt$, as the difference between incident stellar energy, 
$S$, and outgoing infrared radiative flux, $F_{IR}$. This gives the zero-dimensional EBM equation,
\begin{equation}
C\frac{dT}{dt}=S\left(1-\alpha\right)-F_{IR},\label{EBM1}
\end{equation}
where  $\alpha$ is the top of atmosphere planetary albedo and $C$ is effective heat capacity per unit area. Eq. (\ref{EBM1}) describes the climate system as a single point, without any representation of meridional or zonal energy transport by atmospheric dynamics. The value of $S$ can remain constant or can change with time to account for the effect of seasons or other orbital variations. Radiative transfer is included through the parameters $F_{IR}$ and $\alpha$, which can take various functional forms depending on the desired complexity and application.

\begin{figure}[ht!]
\centering
\includegraphics[width=1.00\linewidth]{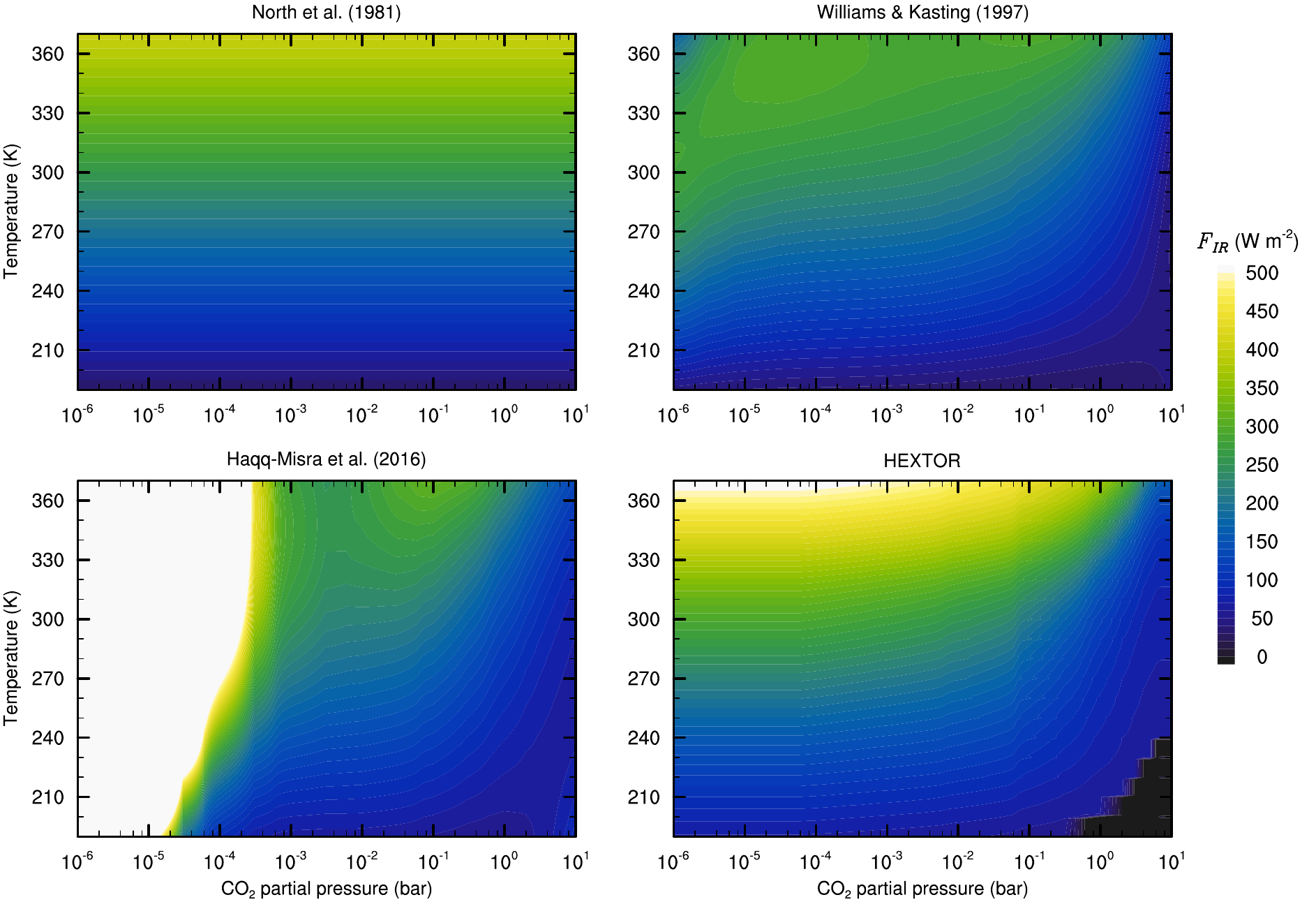}
\caption{Outgoing infrared radiative flux, $F_{IR}$, as a function of temperature and CO$_2$ partial pressure for a linear parameterization (top left, \citet{north1981energy}), two polynomial parameterizations (top right, \citet{williams1997habitable}; bottom left, \citet{haqq2016limit}), and the HEXTOR lookup table method (bottom right). The black region corresponds to numerically unstable solutions where CO$_2$ condenses onto the surface. The lookup table method provides better representation of the underlying radiative-convective equilibrium model solutions for $F_{IR}$, particularly at high temperatures.\label{fig:olrcompare}}
\end{figure}

A common simplifying assumption in EBMs is to assume a linear function for the outgoing infrared radiative flux, $F_{IR}=A+BT$, where $A$ and
$B$ are constants \citep[e.g.][]{budyko1969effect,north1981energy,haqq2014damping}. The value of $\alpha$ is usually either a constant value or a step function that depends on temperature to include
different values for frozen and unfrozen surface. This linear assumption is computationally efficient and provides reasonable accuracy at small departures from Earth's present-day temperature.
However, the linear approximation to $F_{IR}$ tends to underestimate the outgoing infrared radiative flux at high temperatures and also only depends on $T$, with no explicit representation of greenhouse gas abundances such as the carbon dioxide partial pressure, $p$CO$_{2}$ (Fig. \ref{fig:olrcompare}, top left).

One way to improve on a linear function for outgoing infrared radiative flux is to use polynomial parameterizations of $F_{IR}$. \revision{One-dimensional radiative-transfer models represent the atmosphere as a single point and calculate energy transfers through the atmospheric column. Such models have been widely used to calculate the inner and outer edges of the circumstellar habitable zone for the sun and stars of other spectral types, with the \citep{kasting1993habitable} model being one of the primary models that defined this habitable zone.} \citet{williams1997habitable} used the one-dimensional radiative-convective climate model of \citet{kasting1991co2} \revision{and \citet{kasting1993habitable}} to calculate $\sim$300 values of $F_{IR}$ over a broad parameter space of $T$ and $p$CO$_{2}$, which was implemented as a best-fit \revision{fourth-order} polynomial function. \revision{The planetary habitability model developed by \citet{kasting1993habitable} has since been updated, and} other studies \citep{haqq2016limit,batalha2016climate,hayworth2020warming} likewise developed \revision{fourth-order two-piece} polynomial functions for $F_{IR}$ based on hundreds or thousands of calculations with the \revision{improved} radiative-convective model of \citet{kopparapu2013habitable}, which included additional representation of collision-induced absorption in dense CO$_2$ atmospheres. \revised{Other approaches include assuming a function of the form $F_{IR}=\sigma\prime T^4$, where $\sigma\prime$ is a tuning parameter that is can be fitted from observations or GCM calculations \citep[e.g.][]{okuya2019effects}.} Polynomial representations like these are computationally efficient and allow for more realistic exploration of changes in climate to different forcing scenarios. However, polynomial interpolations can also introduce numerical artifacts that limit the range of parameter space that can be reliably explored with the EBM; the limitations of this method include underestimates of $F_{IR}$ at high temperatures \revision{with the \citet{williams1997habitable} parameterization} (Fig. \ref{fig:olrcompare}, top right) and the presence of numerically unstable values of $F_{IR}$ at low $p$CO$_{2}$ that are beyond the bounds of the parameterization by \citet{haqq2016limit} (Fig. \ref{fig:olrcompare}, bottom left). 

\begin{figure}[ht!]
\centering
\includegraphics[width=1.00\linewidth]{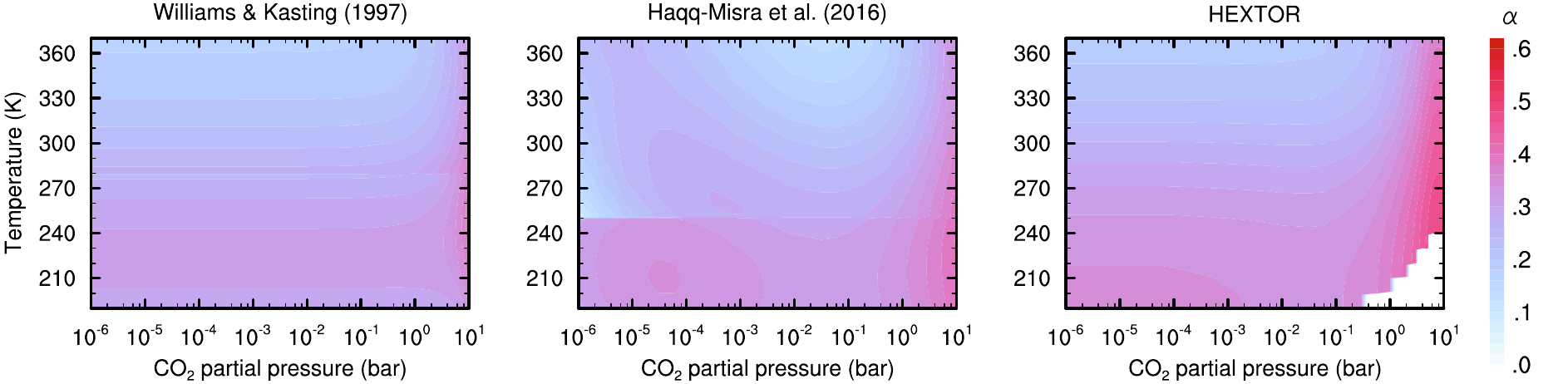}
\caption{Top of atmosphere planetary albedo, $\alpha$, as a function of temperature and CO$_2$ partial pressure, and at a fixed $60^{\circ}$ zenith angle and fixed surface albedo of 0.3, for two polynomial parameterizations (left, \citet{williams1997habitable}; middle, \citet{haqq2016limit}) and the HEXTOR lookup table method (right). Calculations assume a solar spectrum. The white region corresponds to numerically unstable solutions where CO$_2$ condenses onto the surface. The lookup table method provides better representation of the underlying radiative-convective equilibrium model solutions for $\alpha$, particularly at high CO$_2$ partial pressure.\label{fig:albcompare}}
\end{figure}

Polynomial parameterizations can also improve the accuracy of a constant-value assumption of $\alpha$ by allowing the top of atmosphere planetary albedo to depend on the surface type, temperature, and atmospheric composition. \citet{williams1997habitable} used the \revision{radiative-convective climate} model of \citet{kasting1991co2} to calculate 24,000 values of $\alpha$ as a \revision{second-order two-piece} function of $T$, $p$CO$_{2}$, zenith angle, and surface albedo \revision{(which includes the effect of clouds)}. Other studies \citep{haqq2016limit,batalha2016climate,hayworth2020warming} developed similar \revision{third-order two-piece} parameterizations of $\alpha$ using the \revision{improved} model of \citet{kopparapu2013habitable}. These polynomial expressions generally show a high values of $\alpha$ for cold planets and lower values of $\alpha$ for warm planets, while dense CO$_2$ atmospheres also show higher values of $\alpha$ due to increased Rayleigh scattering (Fig. \ref{fig:albcompare}, left and middle). As with any polynomial fit, these expressions for $\alpha$ include some discontinuities and artifacts that do not accurately describe the underlying radiative-convective calculations. 

Another way to ensure accurate representation of radiative transfer in an EBM across a full parameter space is to use a lookup table for $F_{IR}$ and $\alpha$. Lookup tables are populated with the results of radiative-convective calculations for $F_{IR}$ and $\alpha$ at specific points in the parameter space, which improves upon a polynomial parameterization by eliminating any discontinuties or errors that result from overfitting a multi-variate data set. HEXTOR includes such a lookup table for radiative transfer that uses the \revision{radiative-convective climate} model of \citet{kopparapu2013habitable} \revision{to populate the table with values $F_{IR}$ of and $\alpha$}. The HEXTOR lookup table includes 1,748  values of $F_{IR}$ stored in the table as a function of $T$ and $p$CO$_{2}$ and 34,960 values of $\alpha$ stored as a function of $T$, $p$CO$_{2}$, zenith angle, and surface albedo. At each time step, the values of $F_{IR}$ and $\alpha$ are determined by querying the lookup table and interpolating between the closest stored values that match the model conditions of $T$, $p$CO$_{2}$, zenith angle, and surface albedo. 

This \revision{lookup table} method provides a much more complete representation of the underlying radiative-convective equilibrium model's calculations of $F_{IR}$ as a function of $T$ and $p$CO$_{2}$, which includes higher values of $F_{IR}$ at high $T$ and $p$CO$_{2}$ compared to a polynomial approach (Fig. \ref{fig:olrcompare}, bottom right). The lookup table representation of $\alpha$ captures the decrease in albedo with increasing $T$ and also more accurately represents the increase in planetary albedo due to Rayleigh scattering for dense CO$_2$ atmospheres (Fig. \ref{fig:albcompare}, right). The lookup tables for $F_{IR}$ and $\alpha$ also denote a region of instability due to CO$_2$ condensation at high $p$CO$_{2}$ and low $T$. \revision{The present version of HEXTOR halts execution upon reaching this region of CO$_2$ condensation instability, so further model development would be required to explore atmospheres near the point of \finrevision{atmospheric} collapse or freeze-out.} Also note that the lookup table method gives higher values of $F_{IR}$ for hot atmospheres with low $p$CO$_{2}$, which is captured to an extent by the linear approximation but not present in the polynomial fits. One limitation of the lookup table method is that it is more computationally intensive \revision{than a polynomial function}, with a computational complexity proportional to the number of elements in the table. However, this added computational burden is still faster than explicitly coupling a radiative-convective climate model to an EBM. (See Appendix \ref{appendix:lookup} for a complete description of the HEXTOR lookup table method.)

\section{Rapidly Rotating Earth-sized Planets}\label{sec:rapid}

Extending Eq. (\ref{EBM1}) to a one-dimensional model requires a term that represents energy transport from warm to cool areas on the planet. 
For a rapidly rotating planet \revision{with low obliquity}, this thermal gradient occurs from equator to pole and induces the meridional Hadley circulation.  
This poleward energy per unit time leaving a given latitudinal band can be represented as a diffusive process based on the divergence of this thermal energy flux, which allows the one-dimensional EBM equation to be written as 
\begin{equation}
C\frac{\partial T}{\partial t}=S\left(1-\alpha\right)-F_{IR}+\frac{\partial}{\partial x}\left[D\left(1-x^{2}\right)\frac{\partial T}{\partial x}\right],\label{EBMfull}
\end{equation}
where $x$ is the sine of latitude and $D$ is a parameter that represents thermal conductivity. (See Appendix \ref{sec:A1} for a derivation of the diffusive energy transport term on the right side of Eq. (\ref{EBMfull}).) This one-dimensional latitudinal EBM is commonplace in Earth system modeling and can be useful for investigating the latitudinal distribution of temperature for rapidly rotating terrestrial planets \citep[e.g.][]{north1981energy,north2017energy}. 

HEXTOR implements Eq. (\ref{EBMfull}) using \revision{a fixed grid of} 18 latitudinal bands \revision{spaced 10$^{\circ}$ apart}, with $S$ as a function of time and zenith angle to represent the seasonal cycle \citep{williams1997habitable}. Each latitudinal band is initialized with a starting temperature and assigned a geography based on the fraction of land versus ocean. The albedo of frozen surface is set to $0.663$ \citep{caldeira1992susceptibility}. The albedo of unfrozen land is set to $0.3$, while the albedo of unfrozen ocean is a function (based on Fresnel's reflection equations) that depends on zenith angle \citep{williams1997habitable}. The total surface albedo is the weighted sum of ice, land, and ocean at each latitudinal band; this total surface albedo is a functional input to determine the top-of-atmosphere planetary albedo $\alpha$. The heat capacity over land is $C_{land}=5.25\times10^6$\,J\,m$^{-2}$\,K$^{-1}$, the heat capacity over ocean is $C_{ocean}=40C_{land}$, and the heat capacity over ice is $C_{ice}=2C_{land}$ \citep{fairen2012reduced}. The total value of $C$ is the weighted sum of $C_{ice}$, $C_{land}$, and $C_{ocean}$ at each latitudinal band. \revision{For simulations with all land and no ocean, the heat capacity is set to a fixed value $C=5.25\times10^8$\,J\,m$^{-2}$\,K$^{-1}$ everywhere in order to prevent numerical instabilities at the initialization of the model.} The diffusive parameter is set to a constant $D=0.38$\,W\,m$^{-2}$\,K$^{-1}$ \citep{haqq2014damping}. (This is the same value of from the variable value of $D$ calculated by \citet{williams1997habitable} for a 1-bar atmosphere with present-day composition and rotation rate.) HEXTOR also reduces the value of $F_{IR}$ by a fixed amount, $F_{cloud}=8.0$\,W\,m$^{-2}$, which \revision{is physically motivated as a representation of the reduction of outgoing infrared radiation by} clouds \revision{but is used} as a tuning parameter to allow the model to produce a $288$\,K global average temperature for present-day Earth. \revision{(This cloud variable was also used as a tuning parameter by \citet{williams1997habitable}, who used a value of $F_{cloud}=14$\,W\,m$^{-2}$.)} The model finite difference scheme uses a timestep of $\Delta t=12$\,hr for calculating the change in temperature as a function of latitude for a complete orbital cycle. 

\begin{figure}[ht!]
\centering
\includegraphics[width=1.00\linewidth]{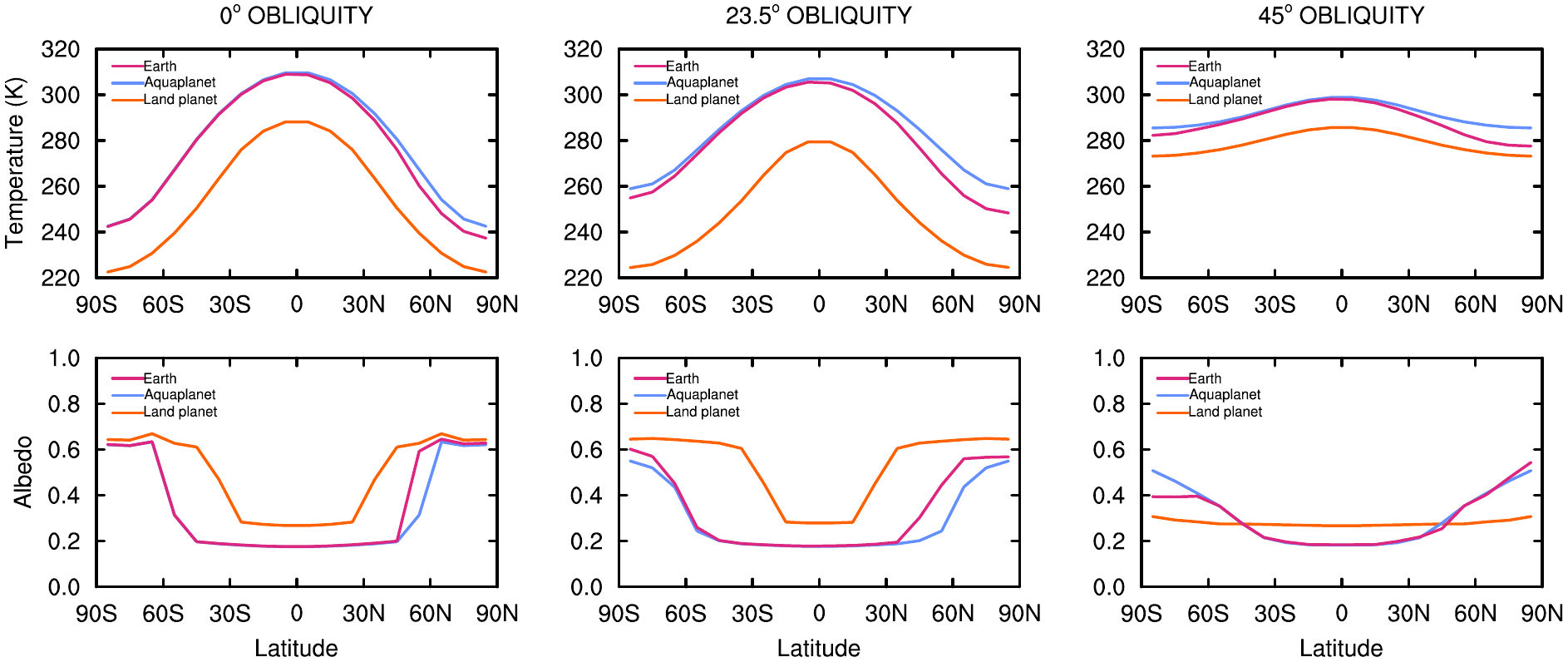}
\caption{HEXTOR calculations of average temperature ($T$, top row) and planetary albedo ($\alpha$, bottom row) for an Earth-sized planet orbiting the Sun with 1\,bar N$_2$ and 400\,ppm CO$_2$. Cases are shown at $0^{\circ}$ (left column), $23.5^{\circ}$ (middle column), and $45^{\circ}$ (right column) obliquity with present Earth (magenta), aquaplanet (blue), and land planet (orange) geography.\label{fig:earthtempalb}}
\end{figure}

A summary of climate states calculated with HEXTOR for an Earth-sized planet orbiting the Sun (with the mean value of $S$ equal to $S_0=1360$\,W\,m$^{-2}$) with 1\,bar N$_2$ and 400\,ppm CO$_2$ is shown in Fig. \ref{fig:earthtempalb}. These calculations all show the converged solutions of $T$ and $\alpha$ after 50 model years, with results at $0^{\circ}$, $23.5^{\circ}$, and $45^{\circ}$ obliquity with present Earth, aquaplanet, and land planet geography. \revision{The Earth geography configuration follows \citet{williams1997habitable} by weighting the surface albedo by the fraction of each latitudinal band covered by ocean versus land.} Global mean values of $T$, $\alpha$, and $F_{IR}$ for these nine cases are shown in Table \ref{table:1}. The latitudinal profiles of $T$ and $\alpha$ are nearly identical for present Earth and aquaplanet geography, with the present Earth cases showing lower temperature and higher albedo near the north pole due to a greater land fraction in the northern hemisphere. The land planet cases show notably lower temperature and higher albedo. (Note that the land planet cases considered in this study still allow for the \revision{formation} of snow on land and \revision{implicitly} assumes a reservoir of groundwater for evaporation, so these cases do not represent arid desert planets.) Increasing the obliquity causes an increase in seasonality that causes the average profiles of $T$ and $\alpha$ profiles to flatten. The most extreme $45^{\circ}$ obliquity calculations all remain ice-free, while the $0^{\circ}$ and $23.5^{\circ}$ obliquity cases all show unfrozen tropics and glacial poles.

\begin{table}[ht!]
\centering
\begin{tabular}{ll|ccc}
Geography     & Obliquity & $T$ (K) & $\alpha$ & $F_{IR}$ (W\,m$^{-2}$) \\
\hline\hline
Present Earth & 0$^{\circ}$         & 287.5           & 0.229  & 263.0                        \\
Present Earth & 23.5$^{\circ}$      & 288.2           & 0.225  & 264.1                        \\
Present Earth & 45$^{\circ}$        & 291.4           & 0.204  & 270.9                        \\
Aquaplanet    & 0$^{\circ}$         & 289.3           & 0.217  & 266.8                        \\
Aquaplanet    & 23.5$^{\circ}$      & 291.8           & 0.201  & 272.1                        \\
Aquaplanet    & 45$^{\circ}$        & 293.4           & 0.190  & 275.3                        \\
Land Planet   & 0$^{\circ}$         & 264.6           & 0.387  & 212.2                         \\
Land Planet   & 23.5$^{\circ}$      & 257.6           & 0.450  & 196.8                        \\
Land Planet   & 45$^{\circ}$        & 280.8           & 0.275  & 246.7                  
\end{tabular}
\caption{Global mean values of temperature, planetary albedo, and outgoing infrared radiative flux calculated by HEXTOR for Earth-sized planets orbiting the Sun with 1\,bar N$_2$ and 400\,ppm CO$_2$.}
\label{table:1}
\end{table}

\begin{figure}[ht!]
\centering
\includegraphics[width=1.00\linewidth]{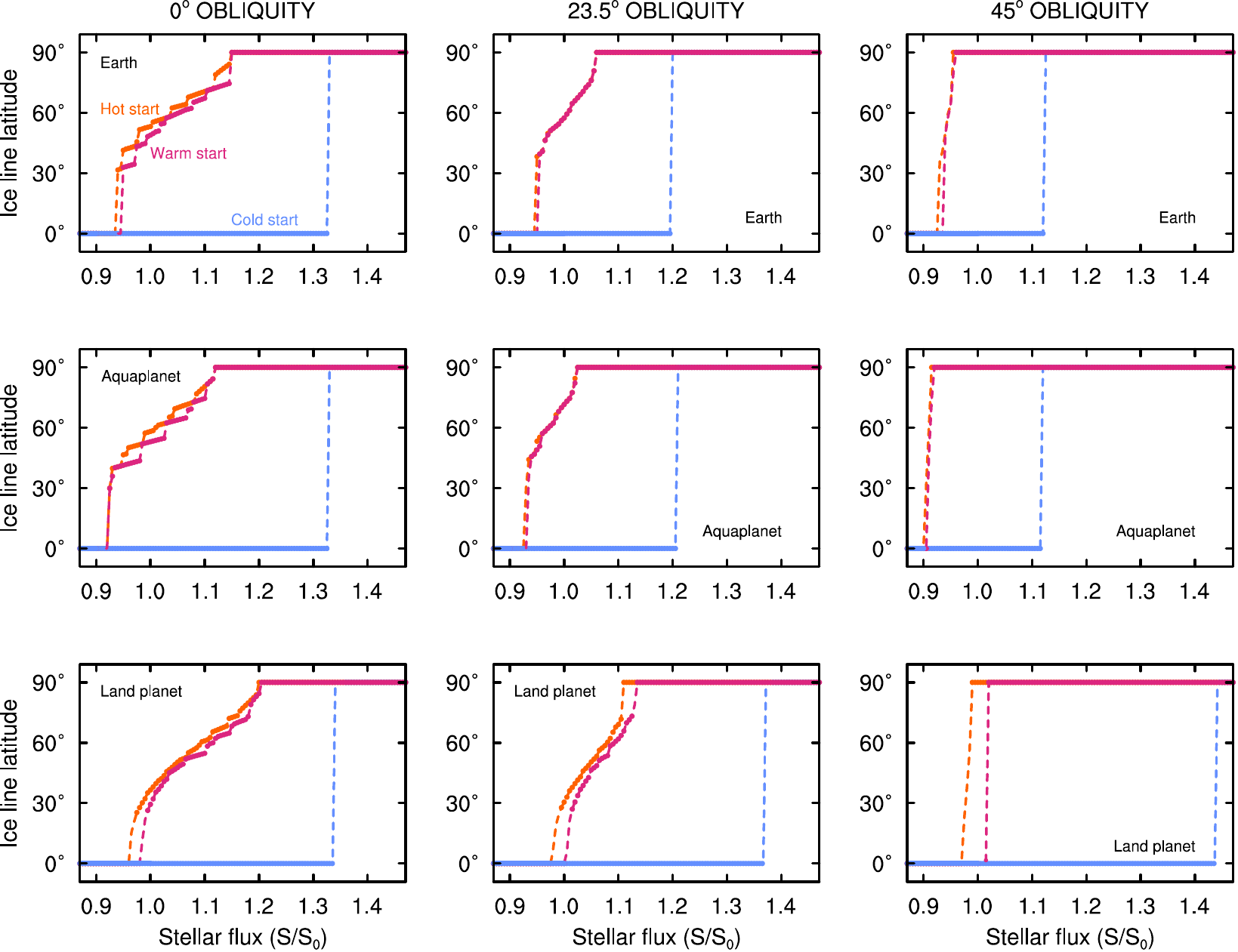}
\caption{Hysteresis in HEXTOR climate states shown as the mean ice line latitude as a function of stellar instellation at $0^{\circ}$ (left column), $23.5^{\circ}$ (middle column), and $45^{\circ}$ (right column) obliquity for present Earth (top row), aquaplanet (middle row), and land planet (bottom row) geography, with $S_0=1360$\,W\,m$^{-2}$. Equilibrium climate states are calculated by initializing the model with \revision{warm ($T=273\,K$, red), hot ($T=300\,K$, orange), and ice-covered ($T=233\,K$, blue) initial} values of $T$. Dashed blue indicates the melting of a frozen planet and dashed red indicates the freezing of a warm planet.\label{fig:earthbistability}}
\end{figure}

Like other EBMs, HEXTOR produces bistable solutions of the equilibrium climate state that depends on the initial values chosen for $T$. Plots of this hysteresis are shown in Fig. \ref{fig:earthbistability} at $0^{\circ}$, $23.5^{\circ}$, and $45^{\circ}$ obliquity with present Earth, aquaplanet, and land planet geography. All calculations assume 1\,bar N$_2$ and 400\,ppm CO$_2$ and show the mean ice line latitude after 50 model years. Three initial values for $T$ are shown that correspond to \revision{warm ($T=273\,K$, red), hot ($T=300\,K$, orange), and ice-covered ($T=233\,K$, blue) initial} climate states. Dashed lines show discontinuous transitions between climate states, with dashed blue indicating the melting of a frozen planet and dashed red indicating the freezing of a warm planet. The hysteresis loop narrows with an increase in obliquity \revision{due to} the enhanced \revision{seasonality}. Aquaplanet cases show a slightly wider hysteresis loop compared to Earth cases, while the land planet cases show \revision{even greater bistability}. Ice caps occur for all $0^{\circ}$ and $23.5^{\circ}$ cases, with a maximum extent of $\sim30^{\circ}$ before the onset of glaciation, which is comparable to results from other EBMs and GCMs \citep[e.g.][]{jenkins1999gcm,hoffman2017snowball,north2017energy}. The $0^{\circ}$ Earth geography and aquaplanet cases show multiple ice cap solutions, \finrevision{which includes a small ice cap instability for} the $23.5^{\circ}$ cases. \revision{(The small ice cap instability does not appear in the previous version of the EBM by \citet{williams1997habitable} unless a linear approximation for $F_{IR}$ is used.) The ``staircase'' feature in the $0^{\circ}$ is partly due to the coarse model resolution and could be smoothed by implementing a finer latitudinal grid \citep[e.g.][]{rose2017ice}.} The $45^{\circ}$ show nearly instantaneous transitions between glacial and ice-free states with only a few stable ice cap solutions. Transitions between climate states on hysteresis plots like these can be driven by changes in stellar flux (increasing or decreasing $S$) as well as by changes in greenhouse warming (such as by increasing or decreasing $p$CO$_{2}$).

\section{Synchronously Rotating Terrestrial Planets}

Synchronously rotating planets pose a challenge to latitudinal EBMs because such planets have one hemisphere fixed toward the star in perpetual daylight and the other side in perpetual night. A latitudinal distribution of temperature cannot provide an adequate description of the mean state of such atmospheres, which are characterized by hot climates beneath the substellar point, freezing temperatures at the antistellar point, and a temperate transition along the terminator regions of the planet. Three dimensional climate models are typically used to explore the temperature distribution between the day and night sides of such planets \citep[e.g.][]{joshi1997simulations,joshi2003climate,merlis2010atmospheric,edson2011atmospheric,carone2014connecting,kumar2016inner,turbet2016habitability,del2019habitable,lefevre20213d}.

One approach to representing the climates of synchronously rotating planets with an EBM is to construct the model so that it \revision{includes} the longitudinal temperature profile of the planet along the equator. An explicit way to do this would be to \revision{derive} the thermal diffusion term in Eq. (\ref{EBMfull}) with a longitudinal dependence, which could be used to construct a two-dimensional EBM with both latitudinal and longitudinal coordinates. Two-dimensional EBMs have been used in Earth system science and have even been applied to the scenario of a planet in synchronous rotation around an M-dwarf star with a binary companion \citep{okuya2019effects}. However, an even simpler approach for obtaining a zonal temperature profile on such planets is to use the same EBM equation as in Eq. (\ref{EBMfull}) but \revision{transform} the coordinate axis so that the $x$ variable \revision{represents the cosine of the tidally-locked latitude, $\theta_{tl}$, where $\theta_{tl}=0^{\circ}$ at the substellar point, $\theta_{tl}=90^{\circ}$ at the terminator, and $\theta_{tl}=180^{\circ}$ at the antistellar point \citep{fortney2010transmission,koll2015deciphering,checlair2017no}. Following \citet{checlair2017no}, the EBM instellation depends on the zenith angle, $\phi_z$, so that instellation is $S\cos\phi_z$ for $\phi_z \leq 90^{\circ}$ and $0$ for $\phi_z > 90^{\circ}$.} \finrevision{This definition makes $\theta_{tl} = \phi_z$ on the day side of the planet.} By making this coordinate transformation and adopting this form of $S$, Eq. (\ref{EBMfull}) can be used to \revision{calculate} temperature on a synchronously rotating planet with a constantly illuminated day side and dark night side.

Following this coordinate transformation approach, the EBM equation (Eq. (\ref{EBMfull})) can be applied to synchronously rotating planets like TRAPPIST-1e. The THAI protocol \citep{fauchez2019trappist} considers four climate configurations for TRAPPIST-1e: two benchmark cases (Ben1 and Ben2) and two habitable cases (Hab1 and Hab2). Ben1 and Ben2 both represent land planets and assume a fixed surface albedo of 0.3, while Hab1 and Hab2 represent aquaplanets with a surface albedo of 0.06 for ocean and 0.25 for ice, \revision{all of which use a constant heat capacity $C=4\times10^7$\,J\,m$^{-2}$\,K$^{-1}$}. Ben1 and Hab1 are both defined with present-Earth atmospheres of 1\,bar N$_2$, 400\,ppm CO$_2$. Ben2 and Hab2 have thicker 1\,bar CO$_2$ atmospheres (although calculations with HEXTOR for the Ben2 and Hab2 cases include 1\,bar N$_2$ plus 1\,bar CO$_2$ due to the constraints of the radiative transfer lookup table). The THAI protocol also specifies $S=900$\,W\,m$^{-2}$ and a 2600\,K stellar spectrum for along with constraints on the rotational and orbital period (6.1\,d), planetary mass (0.772\,$M_\Earth$), and gravitational acceleration (0.930\,$g_{\Earth}$) that are used in the model calculation of orbital cycle and seasons. Because TRAPPIST-1e is assumed to be in synchronous rotation, the model obliquity is fixed at zero. 

\begin{figure}[ht!]
\centering
\includegraphics[width=1.00\linewidth]{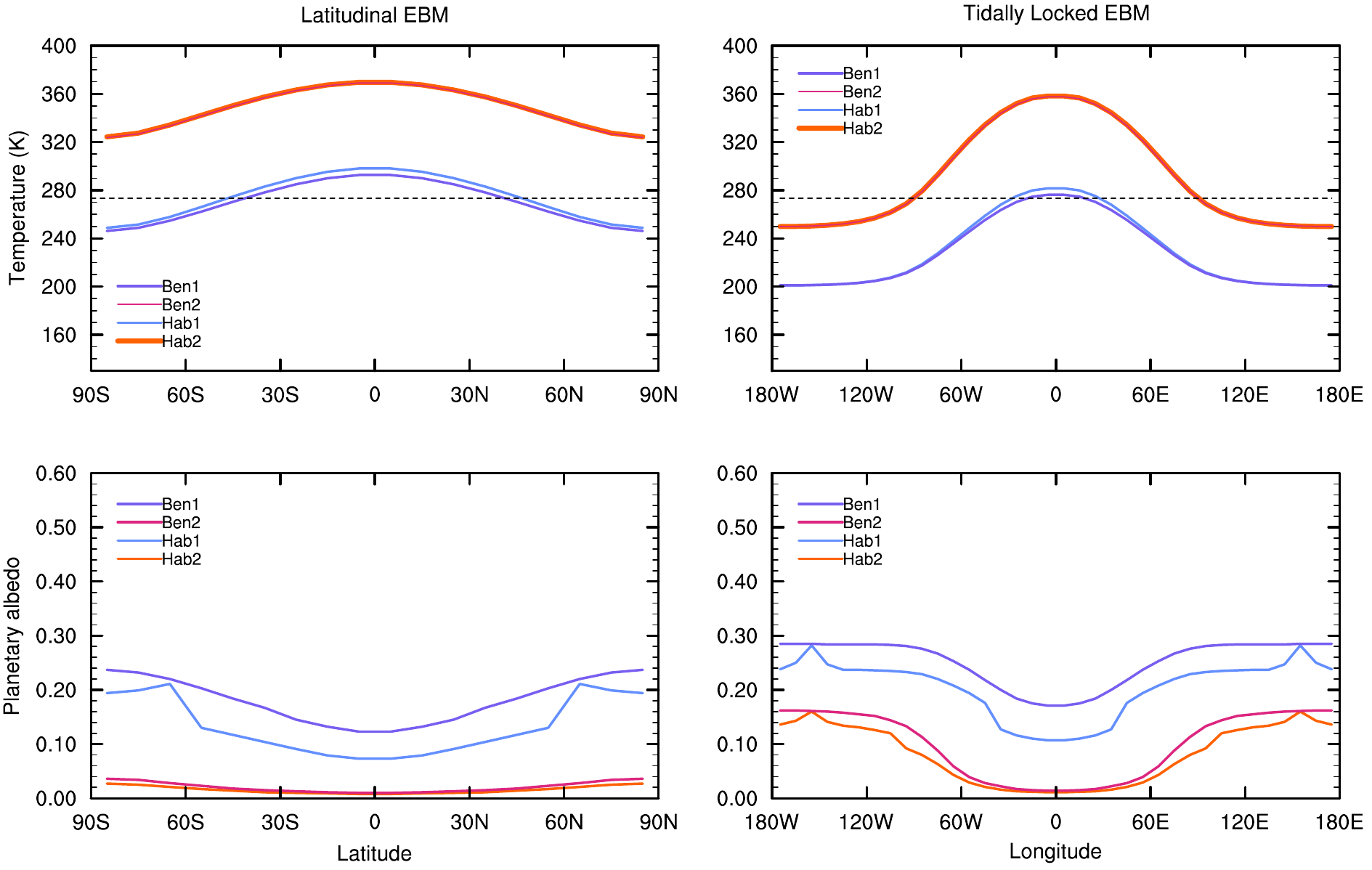}
\caption{HEXTOR calculations of average temperature ($T$, top row) and planetary albedo ($\alpha$, bottom row) for a TRAPPIST-1e-sized planet orbiting the Sun with 1\,bar N$_2$ and 400\,ppm CO$_2$. The left column shows results using a latitudinal EBM, and the right column shows results using a coordinate-transformed \revision{tidally-locked} EBM with maximum flux at the substellar point and zero flux at the night side. Results are shown for the Ben1 (lavender), Ben2 (red), Hab1 (blue), and Hab2 (orange) cases described by the THAI protocol. \label{fig:trappist1s-lat}}
\end{figure}

Calculations with HEXTOR for the four TRAPPIST-1e cases from the THAI protocol are shown in Fig. \ref{fig:trappist1s-lat}, with global mean values of $T$, $\alpha$, and $F_{IR}$ shown in Table \ref{table:1}. The left column of Fig. \ref{fig:trappist1s-lat} shows TRAPPIST-1e calculations using the standard latitudinal EBM with \finrevision{the same functional form of $S$} as the rapidly rotating planets in section \ref{sec:rapid}. These temperature and planetary albedo profiles are markedly flat across latitude, which demonstrates the limitations of directly using a latitudinal model for a synchronously rotating planet. The Ben1 and Hab1 cases are entirely frozen and represent an underestimate of $T$, while the Ben2 and Hab2 cases are completely ice-free and represent an overestimate of $T$. The standard latitudinal EBM is intended for calculating zonal mean $T$ on a rapidly rotating planet like Earth and thus is not ideally suited for representing the day-night contrast between hemispheres of a planet in synchronous rotation.

\begin{table}[ht!]
\centering
\begin{tabular}{llccc}
EBM Configuration     & THAI Case & $T$ (K) & $\alpha$ & $F_{IR}$ (W\,m$^{-2}$) \\
\hline\hline
Latitudinal & Ben1 & 277.9 & 0.153 & 215.4                    \\
Latitudinal & Ben2 & 355.1 & 0.015 & 179.9                      \\
Latitudinal & Hab1 & 282.5 & 0.098 & 221.8                       \\
Latitudinal & Hab2 & 355.6 & 0.011 & 181.1                     \\
\hline
Tidally-locked & Ben1 & 224.5 & 0.218 & 120.7   \\
Tidally-locked & Ben2 & 286.3 & 0.038 & 91.28  \\
Tidally-locked & Hab1 & 226.0 & 0.164 & 123.4   \\
Tidally-locked & Hab2 & 286.7 & 0.028 & 91.76   \\
\end{tabular}
\caption{Global mean values of temperature, planetary albedo, and outgoing infrared radiative flux calculated by HEXTOR in both latitudinal and \revision{tidally-locked} (with hemispheric $S$) modes for TRAPPIST-1e using the four cases from the THAI protocol.}
\label{table:2}
\end{table}

A \revision{tidally-locked} coordinate transformed version of HEXTOR provides a better way to represent this temperature profile in an EBM. \revision{Stellar flux is constructed according to the function by \citet{checlair2017no} described above, which gives an illuminated substellar hemisphere and dark antistellar hemisphere.} The infrared absorption due to clouds is also set to $F_{cloud}=0$ \revision{on both hemispheres, so that the calculated temperature profiles represent the maximum (i.e., clear sky) values predicted by the EBM.} (Note that the cloud infrared effect might \revision{be significant} on the dayside when using the fixed $F_{cloud}=10$\,W\,m$^{-2}$ as the substellar point may be covered in thick tropical clouds driven by strong convection about the substellar point \citep{yang2013stabilizing}.) The resulting longitudinal equatorial profiles of $T$ and $\alpha$ for the four THAI cases are shown in the right column of Fig. \ref{fig:trappist1s-lat}. \finrevision{The tidally-locked EBM temperature profiles have been converted into longitudinal coordinates with the substellar point at $0^{\circ}$, the anti-stellar point at $180^{\circ}$, and the terminator at $90^{\circ}$, with symmetry assumed between east and west hemispheres.} \revision{In contrast to the latitudinal EBM, all four THAI cases now show warm substellar points and cold substellar points when using the tidally-locked EBM.} The land planet cases (Ben1 and Hab1) still have flat albedo profiles from being globally glaciated, but the temperature profiles show a notably colder night side compared to the day side. The ocean planet cases (Ben2 and Hab2) show low albedo on the open ocean day side and high albedo at the glacial night side, with the temperature profile showing the warmest temperatures at the substellar point and coldest temperatures at the antistellar point. \finrevision{The similarity between the Ben2 and Hab2 cases is a result of the dense CO$_2$ atmosphere dominating any albedo contributions from the surface.} The global mean values in Table \ref{table:2} show average temperatures near the freezing point of water for the \revision{tidally-locked} Ben2 and Hab2 cases. 

The \revision{tidally-locked} configuration of HEXTOR shown in the right column of shown in Fig. \ref{fig:trappist1s-lat} provides a qualitative improvement over the standard configuration show in the left column. The \revision{tidally-locked} configuration captures a pronounced day-night temperature contrast as expected for planets in synchronous rotation, \revision{which} show \revision{reasonable consistency with} the predictions of the THAI GCM ensemble. For Ben1, the GCMs find a substellar temperature exceeding 300\,K with a mean temperature of about 215\,K and a night side near 150\,K \citep{turbet2021}, while the \revision{tidally-locked} EBM finds a \revision{slightly colder day side and warmer night side}. The GCM ensemble results for Ben2 show a warm substellar point of about the same temperature as the EBM predicts, but the GCM average finds a colder night side temperature of about 180\,K. For Hab1, the GCMs find a maximum temperature of about 290\,K at the substellar point, \revision{while the} EBM predicts \revision{slightly colder day side temperatures}. The Hab2 case shows the greatest similarity between the two approaches, with the GCM showing a substellar temperature around 320\,K and a colder night side around 265\,K \citep{sergeev2021}; the EBM shows \revision{warmer day side but similar night side temperatures}. 

\begin{figure}[ht!]
\centering
\includegraphics[width=1.00\linewidth]{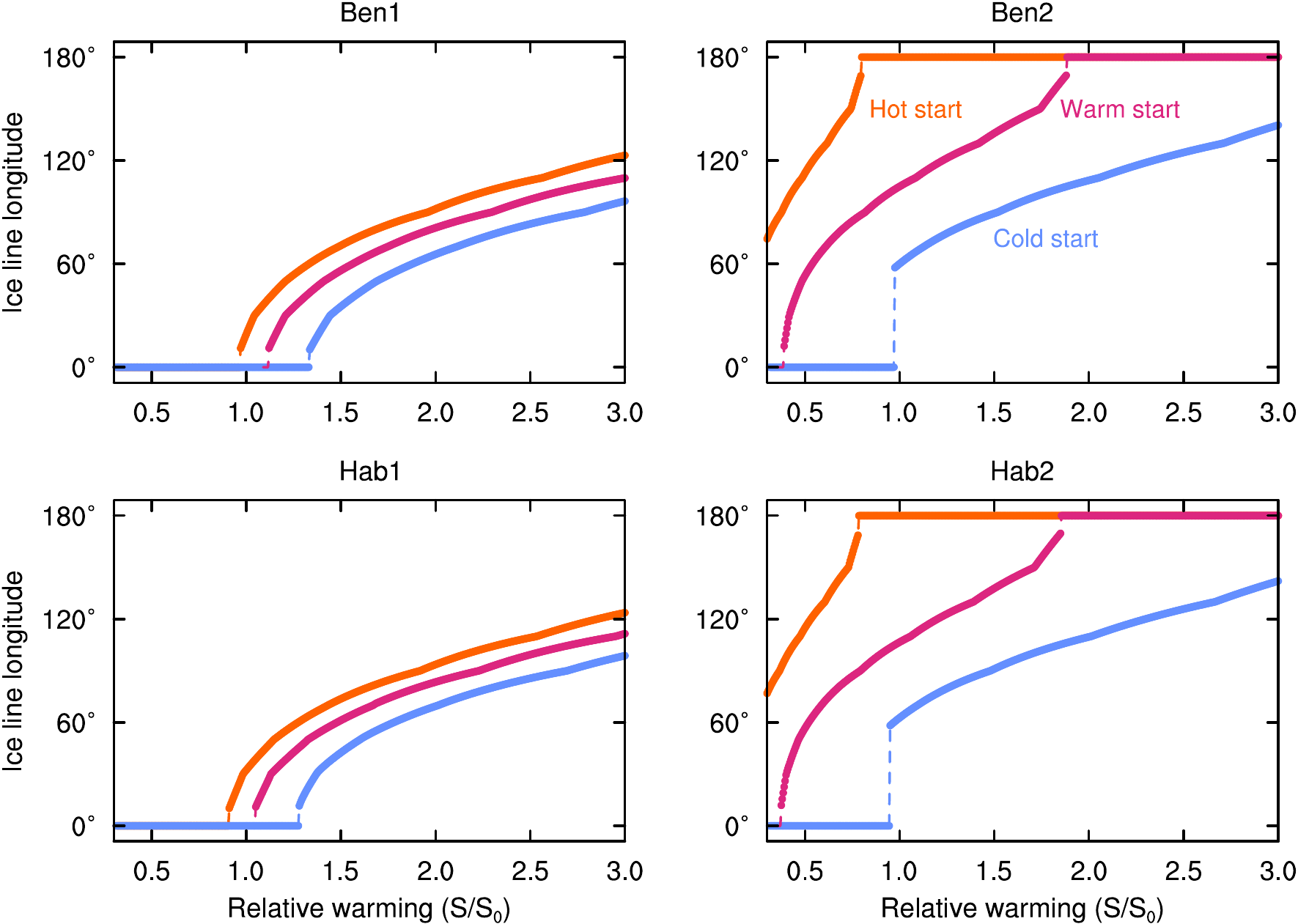}
\caption{HEXTOR climate states for TRAPPIST-1e do not show \revision{strong bistability between cold and ice-free conditions.} Calculations with the coordinate transformed \finrevision{tidally-locked} EBM are shown as the mean ice line longitude as a function of stellar instellation for the Ben1 (top left), Ben2 (top right), Hab1 (bottom left), and Hab2 (bottom right) THAI protocol cases, with $S_0=900$\,W\,m$^{-2}$. Equilibrium climate states are calculated by initializing the model with \revision{warm ($T=273\,K$, red), hot ($T=300\,K$, orange), and ice-covered ($T=233\,K$, blue) initial} values of $T$. Dashed blue indicates the melting of a frozen planet and dashed red indicates the freezing of a warm planet. \revision{The moderate amount of hysteresis that depends on initial conditions is likely due to limitations of the model configuration.}\label{fig:THAIbistability}}
\end{figure}

The behavior of the ice line longitude with changes in $S$ for these THAI cases is shown in Fig. \ref{fig:THAIbistability}. M-dwarf stars like TRAPPIST-1 emit a greater fraction of radiation at longer wavelengths compared to the sun, which leads to a reduction in ice-albedo feedback for planets around such stars \citep{shields2013effect}. The reduction of this ice-albedo feedback leads to the loss of bistability in the climate system, which has been demonstrated using calculations with EBMs and GCMs \citep{checlair2017no,checlair2019no}. Calculations with the \finrevision{tidally-locked} configuration of HEXTOR likewise do not show \revision{bistability between frozen and ice-free states analogous to the rapidly rotating cases shown in Fig. \ref{fig:earthbistability}} but instead show steady transitions between frozen and ice-free states. The presence of partial ice coverage between frozen and ice-free states corresponds to ``eyeball'' climates with warm substellar points and frozen night sides. 

\revision{HEXTOR calculations do show a moderate amount of hysteresis that depends on the initialization state of the model. A similar result was observed by \citet{checlair2017no}, who observed a small amount of hysteresis using a GCM of intermediate complexity when examining synchronously rotating planets with a cold start versus a warm start. Such hysteresis is likely due to limitations of the GCM, and this hysteresis indeed vanished in subsequent experiments by \citet{checlair2019no} with a more complex GCM that included a dynamic ocean. Similarly, this hysteresis in HEXTOR for synchronously rotating planets is likely a model artefact, rather than a physical result, due to the limitations of the diffusive energy transport parameterization. Another complication with simulating these THAI cases is that HEXTOR as currently developed is incapable of representing pure CO$_2$ atmospheres, as all values in the lookup table assume a background atmosphere of 1\,bar N$_2$. The Ben2 and Hab2 THAI cases stipulate a 1\,bar CO$_2$ atmosphere with no N$_2$, so these calculations with HEXTOR will \finrevision{overestimate} the extent of Rayleigh scattering in these cases. This \finrevision{overestimation} of scattering may be another factor in the moderate hysteresis observed for Ben2 and Hab2 in Fig. \ref{fig:THAIbistability}. Future model development could allow HEXTOR to represent pure CO$_2$ atmospheres by constructing a new lookup table.}

The benefit of these comparisons is to examine the extent to which simplified models like EBMs are able to capture the behavior of more complex models, particularly for cases like synchronous rotation. An important source of disagreement between GCM and \revision{tidally-locked} EBM predictions is the representation of day-night energy transport with the parameter $D$ in the EBM. All calculations in this paper have assumed a fixed value of $D$, but $D$ can also be constructed as a spatial (or temporal) variable. \citet{williams1997habitable} allowed $D$ to vary with rotation rate, \revision{mass, gravity, and heat capacity} for a latitudinal EBM; however, implementing this dependence for the THAI cases in a \finrevision{coordinate transformed tidally-locked} EBM led to very large values of $D$ and completely flat profiles of $T$. One way to improve upon this would be to obtain longitudinal profiles of $D$ from a GCM for implementation in an EBM, which would allow for a more accurate parameterization of day-to-night side dynamic transport using the EBM thermal diffusion term. Another improvement could involve the use of scaling factors for thermal energy redistribution that have already been benchmarked against GCM simulations \citep{koll2015deciphering,koll2016temperature,koll2019scaling}. These strategies could be applied to a one-dimensional  \revision{tidally-locked} EBM as well as a two-dimensional EBM that includes both latitude and longitude. Tuning an EBM with a GCM for the spatial distribution of parameters like $S$ and $D$ would have to be done separately for each case of interest, but it would result in greater consistency between the EBM and GCM configurations. 

\section{Conclusions}

HEXTOR is an EBM that has been validated for present-day Earth and can be used to investigate the steady-state and time evolution of climate for rapidly rotating terrestrial planets. This version of the EBM includes a newly-developed lookup table method for calculating the outgoing infrared radiative flux and planetary albedo, which increases the range of temperature and CO$_2$ partial pressures for the model. The lookup table method has an increased computational burden but avoids the spurious behavior that can occur when high-order polynomial fits are used instead. 

One-dimensional EBMs generally are used to calculate latitudinal profiles of temperature, but this configuration is inadequate for calculating the day-night temperature contrast of planets in synchronous rotation around low mass stars. HEXTOR also includes a \revision{tidally-locked} coordinate transformed mode that can be used to calculate the longitudinal equatorial profile of temperature for a planet in synchronous rotation. This  \revision{tidally-locked} mode was used to examine the four cases from the THAI protocol, which provides better one-dimensional representation of a warm day side and cold night side. The consistency of these EBM calculations with GCM results could be improved by using GCMs to constrain the thermal diffusion conductivity parameter as a function of longitude. Other methods of tuning a one-dimensional model using GCM results may be useful for applying EBMs to planets in synchronous rotation and other exotic configurations.

Future advances in modeling exoplanet climates would benefit from an intercomparision focused on EBMs. The primary focus of such an intercomparision should focus on examining differences in the methods used by each EBM for calculating or parameterizing the outgoing infrared radiative flux and planetary albedo. EBM implementations may also show differences in ice-albedo feedback due to different assumptions about the choice of surface properties, but such differences can be minimized by choosing an intercomparision protocol that specifies surface properties. Two-dimensional and other hybrid EBMs would also be useful in such an intercomparision, specifically for understanding any differences in the representation of thermal diffusion in such models. Rapidly rotating planets are the best suited for such an intercomparison, although the continuing discovery and characterization of planets in M-dwarf systems will also continue to motivate development of simplified models to represent these synchronously rotating planets. 

\acknowledgments
J.H.M. thanks Ravi Kopparapu, Eric Wolf, and Thomas Fauchez for helpful comments and conversations and gratefully acknowledges funding from the NASA Habitable Worlds program under award 80NSSC20K0230. This study was conducted as part of the TRAPPIST Habitable Atmosphere Intercomparison (THAI) Workshop, which was hosted by the NASA Nexus for Exoplanet System Science (NExSS) on 14-16 September 2020. This work was facilitated through the use of advanced computational, storage, and networking infrastructure provided by the Hyak supercomputer system at the University of Washington. Any opinions, findings, and conclusions or recommendations expressed in this material are those of the authors and do not necessarily reflect the views of their employers or NASA.

\software{The HEXTOR source code is freely available at \url{https://github.com/BlueMarbleSpace/hextor/releases}. GNU Parallel \citep{tangeole_2020} was used to perform the computations, and the NCAR Command Language \citep{brown2012ncar} and CET Perceptually Uniform Colour Maps \citep{kovesi2015good} were used in post-processing.}

\appendix

\section{Interpolating Radiative Transfer with a Lookup Table}\label{appendix:lookup}
This appendix describes the method used in HEXTOR for interpolating radiative transfer variables from a lookup table. The first step was generating a lookup table of data showing $F_{IR}$ and $\alpha$ as a function of surface pressure ($P$), CO$_{2}$ mixing ratio ($f$CO$_2$), surface temperature ($T$), surface albedo ($a_s$), and zenith angle ($Z$), using the radiative-convective equilibrium (RCE) model of \citet{kopparapu2013habitable}. (Note that the stratospheric temperature in these radiative-convective calculations is fixed following the explanation provided in the Appendix and Equation (6) by \citet{haqq2016limit}.) Lookup tables were calculated for both a sun-like (5800 K) and M-dwarf (2600 K) host star, with the assumption that the atmosphere includes a background N$_2$ pressure of 1\,bar. The calculated 1,748 values of $F_{IR}$ and 34,960 values of $\alpha$ were stored in each lookup table over the range of \revision{$1$\,bar $\le P\le11$\, bar and $10^{-6}\le f$CO$_2\le0.9$ spaced logarithmically, $190$\,K $\le T \le370$\, K spaced by increments of 10\,K, $0.2 \le a_s \le1$ spaced by increments of 0.2, and $0^{\circ} \le Z \le 90^{\circ}$ spaced by increments of $30^{\circ}$. This grid spacing was developed in order to ensure accuracy with the values of $F_{IR}$ and $\alpha$ while maintaining limits to the table size. The computational complexity of this method scales with the number of elements in the lookup table, so significantly increasing the grid size would add to the computational requirements of the model.}

\revision{The lookup table is implemented in the model FORTRAN code as a hash table, which provides a computationally efficient means of storing and retrieving large sets of data.}   
Let $\mathbf{x}$ represent the set of physical parameters that specify
a value of $F_{IR}$ and $\alpha$. For $F_{IR}$, the relevant parameters are $\mathbf{x}_{IR}=\left\{P,f\text{CO}_2,T\right\}$.
For $\alpha$, the relevant parameters are $\mathbf{x}_{\alpha}=\left\{P,f\text{CO}_2,T,a_s,Z\right\}$.
The interpolation approach is identical for both variables, so this discussion will use the generic variable
$\mathbf{x}$. The \revision{hash} table contains $n$ sets of parameters,
so that $\mathbf{x}_{i}$ represents the $i$th set of elements in
the table. The \revision{hash} table also contains corresponding actual values
of $F_{IR}$ and $\alpha$ from the radiative-convective climate model, which we represent
as $v(\mathbf{x})$.
The objective is to find a function $f(\mathbf{x})$
(where $\mathbf{x}$ is given as input) that interpolates from the
values $\mathbf{x}_{i}$ in the \revision{hash} table to obtain the values of $F_{IR}$ and $\alpha$.

The method first finds the nearest neighbor between the input point $\mathbf{x}$ and the values $\mathbf{x}_{i}$ in
the \revision{hash} table. \revision{The nearest neighbor is denoted as $\mathbf{x}_{1}$. The point closest to $\mathbf{x}_{1}$ in the hash table (known as the second nearest neighbor) is} $\mathbf{x}_{2}$. 
The values of $f(\mathbf{x})$ are then calculated as
\begin{equation}
f(\mathbf{x})=v(\mathbf{x}_{1})+\left|\frac{\mathbf{x}-\mathbf{x}_{1}}{\mathbf{x}_{2}-\mathbf{x}_{1}}\right|\left(v(\mathbf{x}_{2})-v(\mathbf{x}_{1})\right).\label{eq:interpolatingfunc}
\end{equation}
Eq. (\ref{eq:interpolatingfunc}) \revision{uses} the \revision{hash} table to calculate an interpolated value based on an arbitrary input $\mathbf{x}$. This expression is implemented in HEXTOR for the calculation of $F_{IR}$ and $\alpha$ at each time step and each latitudinal band.

\section{Diffusive Energy Transport}\label{sec:A1}

Energy transport in an EBM is represented as a diffusive process that is calculated as the divergence of the thermal flux, following the approach of \citet{north2017energy}. 
Letting $\mathbf{q}$ represent the thermal energy flux, the latitudinal divergence of $\mathbf{q}$ in the direction of the polar vector, $\vartheta$, can be written as
\begin{equation}
\nabla_{\vartheta}\cdot\mathbf{q}=\frac{1}{R\sin\vartheta}\frac{\partial}{\partial\vartheta}\left(q_{\vartheta}\sin\vartheta\right)\label{eq:Adiv1},
\end{equation}
where $R$ is planetary radius. Letting $x=\cos{\vartheta}=\sin{\theta}$, where $\theta$ is latitude, Eq. (\ref{eq:Adiv1}) can be rewritten using the chain rule, 
\begin{equation}
\frac{\partial}{\partial\vartheta}=\frac{\partial x}{\partial\vartheta}\frac{\partial}{\partial x}=-\sqrt{1-x^{2}}\frac{\partial}{\partial x}\label{eq:chain},
\end{equation}
to obtain an expression for the divergence of the thermal energy flux, 
\begin{equation}
\nabla_{\vartheta}\cdot\mathbf{q}=\frac{-1}{R}\frac{\partial}{\partial x}\left(q_{\vartheta}\sqrt{1-x^{2}}\right)\label{eq:Adiv2}.
\end{equation}
The thermal flux is assumed to be a function of the temperature gradient, $\mathbf{q}=-\kappa\nabla T$, where $\kappa$ represents thermal conductivity. 
The thermal flux divergence is then $\nabla\cdot\mathbf{q}=-\nabla\cdot\kappa\nabla T$. 
The temperature gradient in the $\vartheta$ direction can be expanded using the chain rule (Eq. (\ref{eq:chain})),
\begin{equation}
\nabla_{\vartheta}T=\frac{1}{R}\frac{\partial T}{\partial\vartheta}=\frac{-\sqrt{1-x^{2}}}{R}\frac{\partial T}{\partial x}\label{eq:tempgrad}. 
\end{equation}
Letting $D$ represent a diffusive parameter defined as $D=\kappa/R^{2}$, the latitudinal thermal energy divergence from Eq. (\ref{eq:Adiv2}) can therefore be written using Eq. (\ref{eq:tempgrad}) to obtain
\begin{equation}
-\nabla_{\vartheta}\cdot\kappa\nabla_{\vartheta}T=\frac{\partial}{\partial x}\left[D\left(1-x^{2}\right)\frac{\partial T}{\partial x}\right]\label{eq:Afinal}.
\end{equation}
The term on the right side of Eq. (\ref{eq:Afinal}) represents the diffusion of thermal energy in a latitudinal EBM and is included in Eq. (\ref{EBMfull}).
\revision{It is worth noting that some studies \citep[e.g.][]{north1979differences} write Eq. (\ref{eq:Afinal}) using a constant value of the diffusive parameter, $D_0$, which does not depend on $x$ and thus appears outside of the partial derivative. EBM implementations may therefore differ based on whether or not $D$ is permitted to vary with latitude.}


\bibliography{main}{}
\bibliographystyle{aasjournal}



\end{document}